\documentclass{aa}
\usepackage{graphics}
\usepackage{epsfig}
\usepackage{graphicx}
\usepackage{txfonts}
\usepackage{color}
\usepackage{subfigure}
\usepackage{psfrag}
\usepackage{natbib}
\begin{document}
\titlerunning{GRBs and giant lightning discharges}
\title{Gamma-ray bursts and other sources of giant lightning discharges in protoplanetary systems}
\authorrunning{B. McBreen et~al.}
\author{B. McBreen$^{1}$ \and E. Winston$^{1}$  \and S. McBreen$^{2}$ \and L. Hanlon$^{1}$}
\institute{Department of Experimental Physics, University
College, Dublin 4, Ireland. \and Astrophysics Missions Division,
Research Scientific Support Department of ESA, ESTEC, Noordwijk,
The Netherlands.}
 \offprints{B. McBreen,
\email{brian.mcbreen@ucd.ie}}
\date{Received $<$date$>$ / Accepted $<$date$>$}
\abstract{ Lightning in the solar nebula is considered to be one
of the probable sources for producing the chondrules that are
found in meteorites.   Gamma-ray bursts (GRBs) provide a large
flux of $\gamma$-rays that Compton scatter and create a charge
separation in the gas because the electrons are displaced from
the positive ions. The electric field easily exceeds the
breakdown value of $\approx$ 1 V m$^{-1}$ over distances of order
0.1 AU. The energy in a giant lightning discharge exceeds a
terrestrial lightning flash by a factor of $\sim 10^{12}$.  The
predicted post-burst emission of $\gamma$-rays from accretion
into the newly formed black hole or spin-down of the magnetar is
sufficiently intense to cause a lightning storm in the nebula
that lasts for days and is more probable than the GRB because the
radiation is beamed into a larger solid angle. The giant
outbursts from nearby soft gamma-ray repeater sources (SGRs) are
also capable of causing giant lightning discharges. The total
amount of chondrules produced is in reasonable agreement with the
observations of meteorites. Furthermore in the case of GRBs most
chondrules were produced in a few major melting events by nearby
GRBs and lightning occurred at effectively the same time over the
whole nebula, and provide accurate time markers to the formation
of chondrules and evolution of the solar nebula.  This model
provides a reasonable explanation for the delay between the
formation of calcium aluminium inclusions (CAIs) and chondrules.

}
\maketitle
\section{Introduction}
Chondrules are typically millimetre-sized stony spherules that
constitute the major component of most chondritic meteorites that
originate in the region between Mars and Jupiter and which fall to
Earth.  They appear to have crystallised rapidly from free
floating molten or partially molten drops.  The large amount of
chondrules present in chondritic meteorites and their possible
crucial role in planet formation, highlights the importance of
identifying the mechanism responsible for their formation.  The
heat source responsible for melting chondrules remains uncertain
and proposed processes have been recently reviewed by
\citet{boss1:1996} and \citet{rubin1:2000}. One of the earliest
proposals for the formation of chondrules is lightning caused by
turbulence in the solar nebula
\citep{whipple1:1966,cameron2:1966}. Lightning is a widespread
phenomenon that is not restricted to terrestrial rainclouds
\citep{uman1:1987}. Spacecraft have detected lightning on Jupiter
and other planets.  This model is not favoured for chondrule
production for a number of reasons including the limitation to
the build up of a large electrostatic potential by the gas
conductivity, the breakdown potential of the gas and the
available energy source
\citep{desch2:2000,eisenhour1:1995,gibbard1:1997,love1:1995,pilipp2:1998,weidends1:1997}.
It has also been pointed out that, if lightning were the mechanism
responsible for chondrule formation, it would have to operate on
a large scale comparable in size with the whole nebula
\citep{pilipp1:1992}.
\newline \indent Almost all the proposed heat sources are local to the solar
nebula.  One exception is that the precursor grains near the
surface of the nebula were melted when they efficiently absorbed
X-rays and $\gamma$-rays from a nearby GRB \citep{mcbreen1:1999}.
This mechanism can produce a large amount of chondrules in the
nebula ($\sim 30$ Earth masses) but it has a low probability of
occurrence ($< 0.1\%$). X-ray melting of material has recently
been demonstrated in the laboratory for the first time using a
powerful synchrotron \citep{duggan:2002,duggan1:2003}. The model
of the X-ray melting of material by a GRB did not include
lightning caused by Compton scattering of $\gamma$-rays by the
gas in the nebula or the $\gamma$-rays from post-burst emission.
The effect of this new process is to induce giant lightning
discharges over the whole nebula.

\begin{figure}
\begin{center}
\resizebox{7cm}{!}{\includegraphics{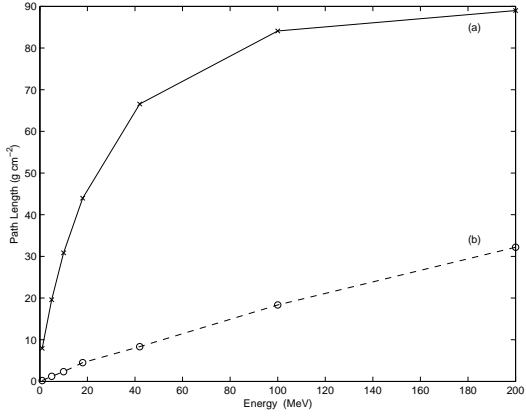}}
\caption{The path length of $\gamma$-rays (a) and range of
electrons (b) as a function of energy in molecular hydrogen. The
$\gamma$-ray data is taken from the compilation of
\cite{cullen1:1989} and the electron data from
\cite{berger1:1964}.}
\end{center}
\end{figure}
\section{Absorption of $\gamma$-rays in the nebula}
The absorption of $\gamma$-rays in a gas with solar abundance is
dominated by molecular hydrogen and it is sufficiently accurate
here to consider only the absorption by H$_{2}$.  Compton
scattering is the dominant process in H$_{2}$ up to 70 MeV above
which pair production and trident production overtake Compton
scattering and only 27\% of $\gamma$-rays with energy of 170 MeV
and 5\% at 1 GeV undergo Compton scattering. For $\gamma$-rays
with energy above a few MeV, the kinematics of Compton scattering
are such that most of the energy is taken by the electron, which
is scattered in the forward direction.  An incoming pulse of
$\gamma$-rays in H$_{2}$ is gradually transformed into electrons
that move further into the nebula leaving a cloud of positive
charge in its wake.  Here we do not consider the scattered
photons that also tend to move in the forward direction.  The
attenuation length of $\gamma$-rays and range of electrons in
H$_{2}$ is given in Fig. 1; the range of the latter in H$_{2}$ is
less than $\gamma$-rays with the same energy.  As the energy
increases, the range of the electrons become an increasing
fraction of the $\gamma$-ray Compton attenuation length, thus
facilitating a larger charge separation.  This process has
similarities to nuclear explosions in the atmosphere
\citep{longmire1:1978}.

A charge separation will also occur with pair production because
some positrons annihilate in flight creating a moving excess of
negative charge that leaves behind a positive excess in the gas
\citep{askaryan1:1962,jelley1:1966}. The cross-section for
positron annihilation is about 1 Barn/$\Gamma$ where $\Gamma$ =
E/m$_{0}$c$^{2}$.  About 10\% of the positrons with energy E =
400 MeV will annihilate in flight.
\begin{figure}
\resizebox{\hsize}{!}{\includegraphics[clip]{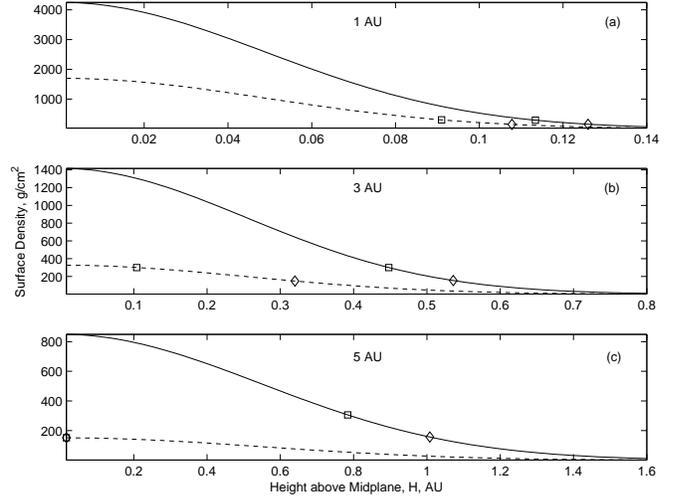}}
\caption{The
surface density, $\Sigma$, as a function of height above the
midplane, H, in astronomical units (AUs) for two models with
normalising values, $\Sigma_{0}$ and exponent n, with
$\Sigma_{0}$ = 4250 g cm$^{-2}$ and n = -1.0 (solid line) and
$\Sigma_{0}$ = 1700 g cm$^{-2}$ and n = -1.5 (dashed line)and for
three radial distances of (a) 1 AU, (b) 3 AU and (c) 5 AU. The
diamonds and squares give path lengths of 150 and 300 g cm$^{-2}$
measured from the top of the nebula.}
\end{figure}
In order to make progress with charge separation it is necessary
to adopt a model of the nebula.  There is a reasonably well
developed theory of the solar nebula including the formation of
terrestrial and gas giant planets.  The discoveries of extrasolar
planets and disks orbiting young stars are leading to the
characterisation of protoplanetary disks around young stars.  In
most models of the solar nebula, after the early evolution of the
Sun, accretion slows down and the nebula becomes essentially
quiescent.  We adopt the power law relationship for the surface
density $\Sigma$ of the solar nebula as a function of radial
distance {\it r} from the Sun given by
\begin{equation}
\Sigma = \Sigma_{0} [r/1AU]^{n} \label{eq1}
\end{equation}
where $\Sigma_{0}$ has the normalising value of 4250 g cm$^{-2}$
at 1.0 AU and exponent n is -1.0 \citep{cameron1:1995}. The
choice of $\Sigma_{0}$ and n varies between the models of the
solar nebula.  \citet{hayashi1:1985} proposed $\Sigma_{0}$ = 1700
and n = -1.5 in the minimum-mass model of the solar nebula and
these values are consistent with minimum-mass models of
extrasolar nebula \citep{kuchner1:2004}.

In conventional models it is usually assumed that lightning
occurs in the dusty midplane of the nebula
\citep{horanyi1:1995,love1:1995}. However a GRB will
preferentially interact with the outer region of the nebula that
is in the direction of the GRB source. Therefore it is necessary
to model the vertical structure of the nebula to obtain the
surface density perpendicular to the midplane.

In the thin disk approximation the midplane isothermal pressure is
given by \citep{cameron1:1995}
\begin{equation}
{\rm P_{c}} = 0.25 \, \Sigma \, \Omega \, \rm{c}_{s} \label{eq2}
\end{equation}
where c$_{s}$ is the speed of sound and $\Omega$
=(GM/r$^{3})^{\frac{1}{2}}$ is the Keplerian angular velocity and
M is the mass of the Sun.  The vertical structure of the disk is
given by
\begin{equation}
\rho = \rho_{0} \, \rm{exp} (-z^{2}/H^{2})
\label{eq3}
\end{equation}
where z is the vertical distance from the midplane.  $\rho_{0}$
and H are the density at the midplane and scale height of the
nebula and are given by $\rho_{0}$ = 2.44 P$_{c}$(NkT)$^{-1}$ and
H = $\pi$ c$_{s} (2 \Omega)^{-1}$ respectively where N is
Avogadros number and k is the Boltzmann constant.

The surface density profile perpendicular to the midplane is
plotted in Fig. 2
for {\it r} = 1, 3 and 5 AU using the two models.  The surface
density is reasonably constant in the midplane region and drops
by only a factor of 2.7 in the first scale height and by the
larger factor of 20 in the second scale height. Approaching the
nebula from above the midplane, a path length of 300 g cm$^{-2}$
will reach the midplane at 5 AU, and 0.1 AU above the midplane at
3 AU using the minimum mass model (Fig. 2c and Fig. 2b).
\section{Prompt GRB emission and charge separation}
The rate of GRBs when averaged over the Hubble volume is $\sim$ 4
$\times$ 10$^{-7}$ yr$^{-1}$ galaxy$^{-1}$ \citep{zhang1:2003}.
The local GRB rate is smaller due to the drastic decrease in star
formation at low redshifts and a value of 0.25 $\times$ 10$^{-7}$
yr$^{-1}$ galaxy$^{-1}$ is widely used at z = 0.  We adopt a value
of 10$^{-7}$ yr$^{-1}$ galaxy$^{-1}$ for the GRB rate and $\sim
10^{7}$ years for the lifetime of a protoplanetary disk. Hence
there is a probability near 100\% that any protoplanetary system
will be irradiated by a GRB. The isotropic equivalent energy of
GRBs is in the range 5 $\times$ 10$^{51}$ to greater then
10$^{54}$ erg.  A GRB with an output of 10$^{52}$ erg will deliver
$\sim 10^{6}$ erg cm$^{-2}$ to a protoplanetary system that lies
at a distance equal to a galactic radius of 10 kpc. The high
energy emission extends to well above 100 MeV with a spectrum
compatible with an extension of the electron synchrotron component
\citep{dingus1:2001}.  An anomalous radiation component in the
energy band between $\sim$ 1-200 MeV was discovered in GRB 941017
\citep{gonzalez1:2003}. This component varies independently of the
prompt GRB emission from 50 keV - 1 MeV and contains three times
more energy.

To model the charge separation and electric field in the nebula,
we adopt a GRB that gives 10$^{6}$ erg cm$^{-2}$ with 10\% of the
energy between 20 MeV and 100 MeV, which is assumed to be in 65
MeV $\gamma$-rays that yield Compton scattered electrons with
energy of 50 MeV.  The incident flux of $\gamma$-rays is $\approx$
10$^{13}$ photons m$^{-2}$ and is attenuated exponentially in
H$_{2}$ using a mass absorption co-efficient of $\mu$ = 1.33
$\times$ 10$^{-2}$ cm$^{2}$ g$^{-1}$ (Fig. 3a (i)). The positive
charge also declines exponentially in the same way because each
Compton event creates a positive ion. The electrons  travel on
average an additional 10 g cm$^{-2}$ into the nebula including
only the ionization loss (Fig. 3a (ii)).
  The path length is further reduced by a
retarding electric field of 1 V m$^{-1}$ which is typical of the
breakdown value (Fig. 3a (iii)). The charge excess is the
difference between the two distributions (Fig. 3b). The net
positive charge is confined to a layer of thickness 10 g
cm$^{-2}$ whereas the negative charge is distributed over a much
wider range.  The net positive or negative charge is 11\% and has
a value of q $\approx$ 10$^{-7}$ C m$^{-2}$. Two percent of the
net negative charge is beyond 300 g  cm$^{2}$ and penetrates
deeply into the nebula.
\begin{figure}
\resizebox{\hsize}{!}{\includegraphics[clip]{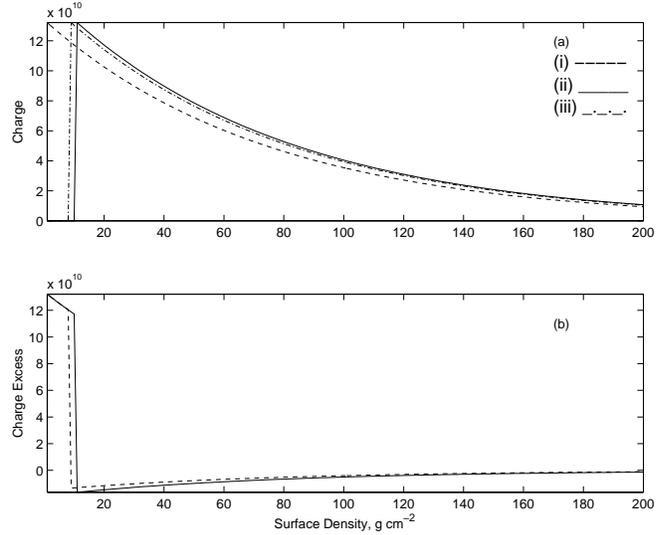}}
\caption{
(a) The number of $\gamma$-rays that Compton scatter and
and the number of positive ions as a function of path length
in H$_{2}$ for an incident flux of 10$^{13}$ photons m$^{-2}$ (i).
The number of electrons as a function of path length taking
into account the ionization loss of 10 g cm$^{-2}$ (ii) and
also including  a retarding electric field of 1 V m$^{-1}$ (iii).
(b) The charge excess as a function of path length 
for the ionization loss (solid line) and also including the retarding
electric field (broken line). The model used is for $r$ = 3 AU and
$\Sigma_{0}$ = 4250 g cm$^{-2}$ with n = -1.0.} 

\end{figure}

Electrical breakdown occurs when a normally insulating gas
suddenly becomes conducting in a strong electric field.  The gas
pressure is particularly important because lower pressure reduces
the voltage necessary for the discharge and increases the width of
the discharge channel.  Breakdown occurs when the electric field
is strong enough that a free electron accumulates $\sim$1 eV of
energy between successive collisions with gas molecules
\citep{pilipp1:1992}. This condition is given by
eE$_{s}$(n$\sigma)^{-1} \approx$ 1 eV where E$_{s}$ is the
discharge electric field, n is the number density of gas
molecules and $\sigma \sim$ 10$^{-15}$ cm$^{-2}$ is the collision
cross section.  E$_{s}$ has a value of $\sim$ 2.4 $\times 10^{6}$
V m$^{-1}$ for air and $\sim$ 2.0 $\times 10^{6}$ V m$^{-1}$ for
H$_{2}$ at atmospheric pressure \citep{love1:1995}. The measured
value of E$_{s}$ in thunderstorms is about a factor of 10 lower
and this difference is often attributed to energetic runaway
electrons from cosmic rays or radionuclides that prematurely
trigger the discharge \citep{gurevich1:2001}.  The value of
E$_{s}$ scales with pressure and has a value of 20 V m$^{-1}$ to
1 V m$^{-1}$ at pressures between 10$^{-5}$ and 5 $\times$
10$^{-7}$ of atmospheric pressure, typical of disks in planetary
forming systems. E$_{s}$ could have a higher value in dust loaded
regions near the midplane \citep{gibbard1:1997}. The Compton
scattered electrons produce ions and electrons that increase the
conductivity of the gas above that caused by cosmic rays and
radionuclides (Desch \& Cuzzi 2000, Love et al. 1995). The large
scale and rapid formation of the charge separation prevents
significant discharging by the gas conductivity.

The charge separation described in the nebula is analogous to a
capacitor.  The basic equation for a parallel plate capacitor
yields the potential difference V = qd($\epsilon_{o})^{-1}$
between the plates separated by a distance d. V attains a value of
10$^{14}$ V for q = 10$^{-7}$ C m$^{-2}$ and a representative
value of d = 0.1 AU.  The electric field is $\sim$ 10$^{4}$ V
m$^{-1}$ and greatly exceeds E$_{s}$ by a factor of at least
10$^{3}$ even for a GRB at a distance 10 kpc and $\sim$ 40 for a
similar GRB at the distance to the Large Magellanic Cloud.  As
the GRB interacts with the nebula, the charge separation creates a
strong electric field that exceeds E$_{s}$ and triggers the
discharge \citep{dwyer1:2003}. The duration of the discharge is
estimated at 100 s \citep{pilipp1:1992}. The average current in
the channel is $\sim$ 10$^{11}$ A if half of the excess charge
from an area $\sim$ d$^{2}$ flows down the channel in 100 s. The
width of the lightning channel depends on the gas pressure and
has an estimated value of \(\sim 10^{5}\) cm assuming it is
limited to a few thousand electron mean free paths
\citep{pilipp1:1992}.

The total energy in $\gamma$-rays, over an area comparable in
size to the charge separation of 0.1 AU, gives an upper limit of
10$^{29}$ erg to the energy dissipated in the channel and this
exceeds a large terrestrial lightning flash by $\sim$ 10$^{12}$.
We cannot exclude the possibility of repeated strikes over the
lightning channel, a situation that is somewhat analogous to the
stepped leaders and return strokes in terrestrial lightning.
Furthermore there is the possibility that the lightning may
fragment into many channels \citep{uman1:1987}.  The visible and
ultraviolet radiation from the discharge heats and melts the
precursor grains to form chondrules out to a distance of \(\sim
10^{9}\) cm from the discharge channel.  The total amount of
chondrules produced in the nebula within a radius of 5 AU is 5
$\times$ 10$^{20}$ g assuming a GRB with 10$^{6}$ erg cm$^{-2}$
has 10$^{5}$ erg cm$^{-2}$ in 65 MeV $\gamma$-rays, 2 $\times$
10$^{10}$ erg g$^{-1}$ to heat and melt the precursor dustballs
and an efficiency of 10$^{-2}$ to convert the $\gamma$-ray energy
to chondrules \citep{jones1:2000}. The amount of chondrules
produced is too small to account for the total mass of $\sim$ 3
$\times$ 10$^{24}$ g in the asteroid belt and the value of $\sim$
10$^{27}$ g when the asteroid belt was 300 times larger than at
present. The mass of chondrules can be increased by 3 $\times$
10$^{4}$ to 1.5 $\times$ 10$^{25}$ g for a GRB at a distance of
300 pc ($\times$ 10$^{3}$) with a higher isotropic luminosity of
10$^{53}$ erg ($\times$ 10) and an anomalous MeV component as
observed in GRB 941017 ($\times$ 3). This model is not sufficient
because there is evidence from compound chondrules
\citep{wasson1:1993} and compositional gradients in chondrules
\citep{wasson2:2003} that they were melted several times
requiring a repeating process.
\section{GRB post-burst emission and other sources}
The afterglow continues for days and weeks after the GRB when the
relativistic blast wave decelerates by sweeping up the external
medium.  Anomalies in the afterglow of some GRBs including the
detection of GeV $\gamma$-rays that lasted for over 90 minutes
after GRB 940217 \citep{hurley1:1994} can be accounted for by
post-burst emission. After the GRB there could be a more extended
period in which energy is injected into the remnant. This
activity may be caused by continued drainage of matter left over
from the explosion into the newly formed black hole or the
spin-down of a super-pulsar or magnetar \citep{rees1:2000}. The
$\gamma$-ray post-burst emission has not been measured and we
must rely on model computations \citep{ramirez1:2004}. The
relativistic outflow from a super-pulsar interacts with photons
from a binary companion or from the explosion. The emission is
mostly in $\gamma$-rays and estimated at 10$^{48}$ erg s$^{-1}$
immediately after the GRB that declines to 10$^{47}$ erg s$^{-1}$
after 10 hours and 10$^{45}$ erg s$^{-1}$ after 100 hours.  The
emission is beamed into a large solid angle that is not well
constrained by the model and we estimate it covers 10$^{-1}$ of
the sky. The prompt GRB fireball emission is collimated into a
much smaller angle that is estimated on average to be 2 $\times$
10$^{-3}$ of the sky \citep{frail1:2001}. The nebula is 50 times
more likely to be irradiated by post-burst emission than by the
GRB.

The model of GRB emission is not unique and in the case of a
structured jet, where the energy density per unit solid angle
falls away from the axis, the emission is beamed into a larger
solid angle by a factor of about 5 \citep{zhang1:2003}. In this
case the post-burst emission is ten times more likely to impinge
on the nebula than the GRB. The post-burst emission of up to
10$^{52}$ erg in $\gamma$-rays is sufficiently intense to cause a
major lightning storm in the disk that lasts for days or even
weeks and up to 50 such events may occur over the lifetime of the
disk.  These fifty events, occurring at a distance of 10 kpc,
will produce 5 $\times$ 10$^{23}$ g of chondrules 
assuming an efficiency of 10$^{-2}$ for
conversion of $\gamma$-ray energy to chondrules.  The largest
mass of chondrules is produced by the post-burst emission from
the nearest GRB;
a burst 100 pc
distant will produce 10$^{26}$ g within a radius of 5 AU in a
disk. Strong dependence on the distance between the GRB and the
protoplanetary disk results in a wide range in the mass of
chondrules produced.  In the above estimates we have assumed that
the charge separation is not removed by the conductivity provided
by the ions and electrons produced by the Compton scattered
electrons, cosmic-rays and radionuclides.  This will only be
correct if the post burst emission is in short duration outbursts
like the GRB emission. The post-burst emission is more likely to
be highly variable in the case of continued and sporadic
accretion into the newly formed Kerr black hole (McBreen et al.
2002).

Soft gamma-ray repeaters (SGRs) are also a possible
source of lightning in the nebula.  The energy that drives the
giant flares ($> 10^{44}$ erg) such as the 1979 March 5 event
from SNR N49 may be caused by a sudden large scale arrangement of
the magnetic field which releases magnetic energy
\citep{thompson1:1996}.  The extreme possibility is that the
entire dipole moment is destroyed in a single event releasing
$\sim 10^{47}$ erg \citep{eichler1:2002}.  The rate of such
events could be as high as the rate of magnetar production $\sim
10^{-3}$ yr$^{-1}$ and $\sim 10^{4}$ such events could occur over
the lifetime of the nebula.  About 10 to 100 of these SGR
explosions should be close enough to generate lightning in the
disk and each one produce a small amount of chondrules,  but the
overall contribution of SGRs is much smaller than that of GRBs.

Other nearby and less variable $\gamma$-ray sources such as
quasars, microquasars and powerful x-ray binaries (Mirabel \&
Rodriguez 1999) might be expected to generate charge separation
and lightning. They require a much longer time of $\sim 10^{5}$ s
to generate an electric field that can exceed E$_{s}$, but the
gas conductivity may limit the electric field to below this value.
There is sufficient energy in these sources to produce large
amounts of chondrules provided a method is found to limit the gas
conductivity. The same limitation has been noted in other models
of charge separation and lightning in the protoplanetary disk
(Desch \& Cuzzi 2000, Gibbard et al. 1997, Love et al. 1995).
\section{Discussion}
GRBs and other sources are capable of producing significant
quantities of chondrules in protoplanetary systems anywhere in
the disk of a galaxy.  Stars form in giant molecular clouds of
size about 100 pc and hence there is a higher than average
probability it will contain massive stars that produce GRBs and
SGRs and irradiate protoplanetary systems in the same cloud to
form large amounts of chondrules.

Compound chondrules consist of a primary that solidified first
and one or more secondaries attached to the primary
\citep{wasson1:1993,wasson2:2003}. Sibling compound chondrules
have very similar textures and compositions and most, perhaps
all, consist of chondrules melted in the same event. If this
event is identified with a nearby GRB, the chondrules would have
been produced all across the disk and should provide a
simultaneous time marker between sibling chondrules in different
meteorites. If the chondrules were produced by GRBs, the
differences between the composition of meteorites are due to
compositionally segmented regions of the nebula.  The independent
compound chondrules probably were produced in two separate events
and it may be possible to determine the order and frequency of
the GRB events from the meteorite record.

The magneto-rotational instability (MRI) provides an
understanding of radial mixing and turbulence in the disk
\citep{balbus1:1998}.  Turbulence can concentrate dust particles
of a particular size to spatial densities well above their
background values \citep{cuzzi1:2001,wood1:1997}. A large amount
of chondrules can then be produced by lightning and provide an
explanation for the large quantity of compound chondrules in
meteorites.  The magnetic field in the nebula must be well
coupled to matter for MRI to be effective and this condition was
satisfied close to the Sun and beyond 10 AU where cosmic ray
ionization may suffice to maintain a significant amount of
ionization.  The magnetic field during chondrule
formation is not well constrained by the meteoritic evidence and
seems to have a value between 1 and 10 gauss
\citep{sugiura1:1979,jones1:2000}. A magnetic field of this
magnitude is significant because it would channel Compton
scattered electrons into regions of enhanced magnetic field, such
as MRIs, and possibly cause lightning between regions in the disk
which is analogous to cloud to cloud lightning in the Earth's
atmosphere.

The $^{26}$Al/$^{27}$Al ratio of CAIs in meteorites has a value of
5 $\times$ 10$^{-5}$ whereas in chondrules the ratio has a much
smaller value \citep{jones1:2000,cameron1:1995}. The simplest
explanation is that there was an interval of at least a million
years between the formation of CAIs and chondrules. Chondrules
should have formed preferentially late in the development of the
disk after the  Sun had stripped away most of the gas and allowed
the $\gamma$-rays to penetrate closer to the midplane.  This
effect is more important in models of the disk with more than the
minimum mass as shown in Fig. 2. The penetration of the disk by
$\gamma$-rays is easier beyond the snowline at about 5 AU because
the disk is less massive. Chondrules produced beyond about 5 AU
may have formed earlier and hence could contain $^{26}$Al.

GRBs in the galaxy affect the Earths atmosphere
\citep{thorsett1:1995, KurtZ:1996,scalo1:2002} and these events
may be responsible for mass extinctions
\citep{dar1:1998,melott1:2004,thomas1:2004}. Fortunately the rate
of GRBs is very low in our galaxy.  However a GRB can reveal
protoplanetary systems in other galaxies by transient infrared
emission from the melting chondrules and optical emission from the
gas \citep{duggan1:2003,mcbreen1:1999}.  These events may be
detectable for hundreds of years after the GRB when the expanding
shell or cones of radiation interacts with protoplanetary systems.
\bibliographystyle{aa}
\footnotesize{
}
\end{document}